\documentclass[twocolumn,twoside,slac_two]{revtex4}
\usepackage{graphicx}
\usepackage{nccfancyhdr}
\renewcommand{\headrulewidth}{0pt}
\renewcommand{\footrulewidth}{0pt}

\begin{document}

\title{Status of a DEPFET pixel system for the ILC vertex detector}

\author{M. Trimpl, M. Koch, R. Kohrs, H. Kr\"uger, P. Lodomez, L. Reuen, C. Sandow, E.~v.~T\"orne, J.J. Velthuis, N. Wermes}
\affiliation{Physikalisches Institut, University of Bonn, Nussallee 12, D-53115 Bonn, Germany}
\author{L. Andricek, H.G. Moser, R.H. Richter, G.Lutz}
\affiliation{Max-Planck-Institut f\"ur Physik, F\"ohringer Ring 6, D-80805 M\"unchen, Germany}
\author{F. Giesen, P. Fischer, I. Peric}
\affiliation{Institut f\"ur Technische Informatik, B\,6, 26, D-68131 Mannheim, Germany}
\begin{abstract}
We have developed a prototype system for the ILC vertex detector based on DEPFET pixels.
The system operates a $128\times64$ matrix (with $\rm{\sim35\times25\,\mu m^2}$ large pixels) and uses two dedicated microchips,
the SWITCHER~II chip for matrix steering and the CURO II chip for readout.
The system development has been driven by the final ILC requirements which above all
demand a detector thinned to 50\,$\mu$m and a row wise read out with line rates of 20\,MHz
and more. The targeted noise performance for the DEPFET technology is in the range of
ENC=100\,e$^-$. The functionality of the system has been demonstrated using different
radioactive sources in an energy range from 6 to 40\,keV. In recent test beam
experiments using 6\,GeV electrons, a signal-to-noise ratio of $\rm{S/N\sim120}$ has been
achieved with present sensors being 450\,$\mu$m thick.
For improved DEPFET systems using 50\,$\mu$m thin sensors in future, a signal-to-noise 
of 40 is expected.
\end{abstract}

\maketitle

\thispagestyle{fancy}

\section{Introduction}
The future International Linear \mbox{Collider (ILC)} will enter a new era of vertex detection.
With the aimed impact parameter resolution of $\mathrm{\sigma\,(d_0) = (3.9\,\oplus\,7.8 / p \cdot\sin{^\frac{3}{2}\theta})\,\mu m}$
precision physics will become possible which complements the discovery potential of the LHC machine.
To achieve such outstanding performance, the ILC vertex detector requires a row wise operation 
where all sensor periphery is placed outside the sensitive area and detector layers that are thinned to 50\,$\mu$m or even below.
Furthermore, the emerging background rates due to the high luminosity beam necessitate 
the readout of the whole detector in 50\,$\mu$s which translates into line rates of 20\,MHz and more.
\section{DEPFET technology for the ILC}
\begin{figure}
  \includegraphics[width=80mm]{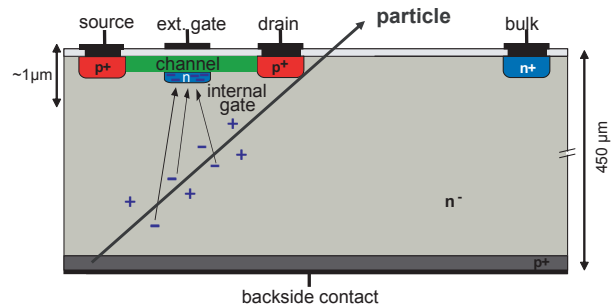}
  \caption{Cross section of a DEPFET pixel illustrating the basic principle of operation.}
  \label{DEPFET_principle}
\end{figure}

One technological option presently discussed for the ILC vertex detector is the DEPFET concept.
The basic principle of operation of the DEPFET is illustrated in figure~\ref{DEPFET_principle}.
The DEPFET (abbreviated \textbf{DEP}leted \textbf{F}ield \textbf{E}ffect \textbf{T}ransistor) \cite{CITE_DEPFET} combines in-pixel amplification and
particle detection by embedding a FET in high-ohmic silicon material.
The sensor is fully depleted due to sidewards depletion from the backside contact and from the transistor channel jointly.
The sidewards depletion and additional implantations close to the detector front side form a 
laterally confined potential valley underneath the transistor channel.
This potential valley is called internal gate, derived from the external gate of the transistor.
Electrons generated by impinging radiation drift to the internal gate where they are collected. 
The potential of the internal gate changes due to the collected charge $\rm{\Delta q}$ and the transistor current is modulated according to $\mathrm{\Delta I_D = g_q \cdot \Delta q}$,
where $\rm{g_q}$ is called the internal gain of the DEPFET transistor, equivalent to the transconductance of the external gate $\rm{g_m}$.
Finally, the signal charge collected in the internal gate can be determined by measuring the device current.  
Since the readout of the device current does not affect the charge in the internal gate, it has to be removed from time to time.
This is know as the clear of the device and is realized by an additional contact at the pixel fringe (not shown in figure~\ref{DEPFET_principle}).
The main advantages of the DEPFET technology are the fully depleted bulk that is used for
signal generation and the fact that charge collection is induced by an electric field.

Since the DEPFET concept relies on a backside contact to provide the fully depletion, 
new thinning concepts apart from the standard ones used in microchip industry need to be explored to fabricate thin DEPFETs.
The present approach for thinning DEPFET modules is based on wafer bonding technology and 
anisotropic etching.
The principle has been demonstrated by producing thin diode structures \cite{CITE_DEPFET_Thinning} and will be combined with a large scale DEPFET pixel production in future.
The radiation tolerance of the sensor against ionizing radiation has been shown up to a dose of 1\,MRad using X-rays from a $\mathrm{^{60}Co}$ source \cite{CITE_LaciRadHard}.

\begin{figure}
  \includegraphics[width=65mm]{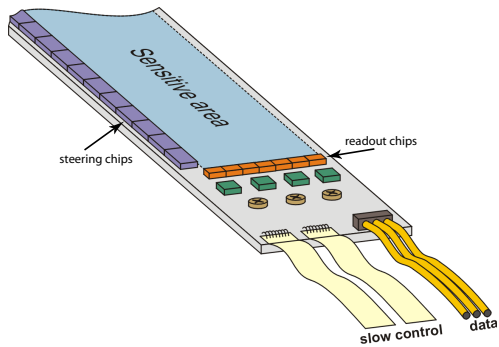}
  \caption{Sketch of one end of an ILC vertex detector laddder with thinned sensitive area, supported by a silicon frame that carries steering and readout chips.}
  \label{ILC_Ladder}
\end{figure}

A principle sketch of one ladder end proposed for a DEPFET ILC vertex detector is shown in figure~\ref{ILC_Ladder}.
The sensitive area is thinned to $50\,\mu$m keeping a thicker frame ($\sim300\,\mu$m) for 
mechanical stability and stiffening of the 100\,mm long and 13\,mm wide ladder.
The steering chips for row selection and clearing strobes are situated at the longer side
so that the detector is read out row wise to the ladder ends.
There, readout chips are placed that process all matrix columns in parallel.

\begin{figure}[b]
  \includegraphics[width=85mm]{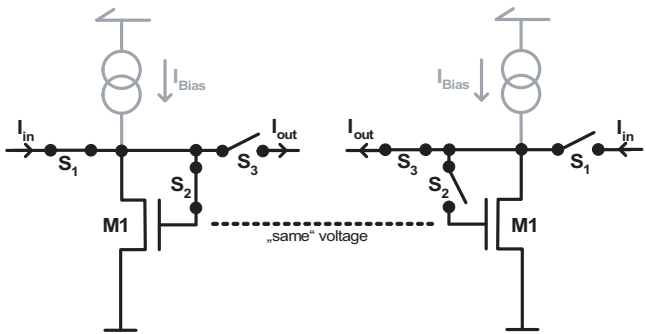}
  \caption{Basic principle of a current memory cell \cite{CITE_SICell} used for signal processing in the readout chip.
  The circuit configuration is shown in the sampling mode (left) and in the hold mode (right).}
  \label{SI_Cell}
\end{figure}

Concerning the readout of such a DEPFET ladder, new concepts have 
been explored as well \cite{CITE_Trimpl_VTX02}.
Since the device signal of the DEPFET is a current, the most natural way is to adapt the
signal processing queue to the current operation of the device.
Therefore, current memory cells are used in the readout chip for temporal storage and processing of the signal current.
The basic principle of operation of such a memory cell \cite{CITE_SICell} is shown in figure~\ref{SI_Cell}.
The cell operates as a dynamic current mirror, where transistor M1 works as the input and output stage depending on the setting of the different switches S1-S3.
Apart from the fact that a direct signal processing without any current to voltage conversion is smart, additional advantages of operating in the current domain exist, upon the most important ones are:
\begin{itemize}
  \item{The voltage sampled at the gate of the transistor M1 in figure~\ref{SI_Cell} is not directly proportional to the signal current.
  Hence, a higher dynamic range compared to a voltage sample and hold can be achieved.
  This is particularly important for emerging process technologies with reduced supply voltage.}
  \item{The subtraction of two values, as needed for pedestal subtraction, can be done very accurate and easily with currents.}
\end{itemize}
\section{The DEPFET pixel system}
\begin{figure}
  \includegraphics[width=80mm]{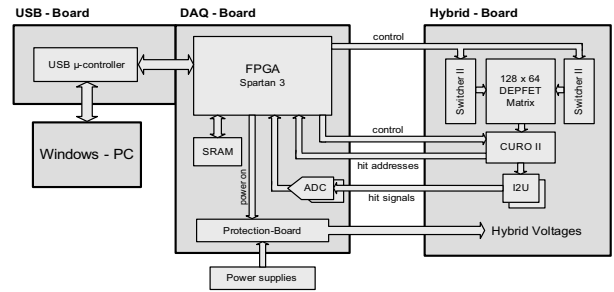}
  \caption{Schematic overview of the ILC DEPFET pixel system.}
  \label{System_overview}
\end{figure}

\begin{figure*}[t]
  \centering
  \includegraphics[width=135mm]{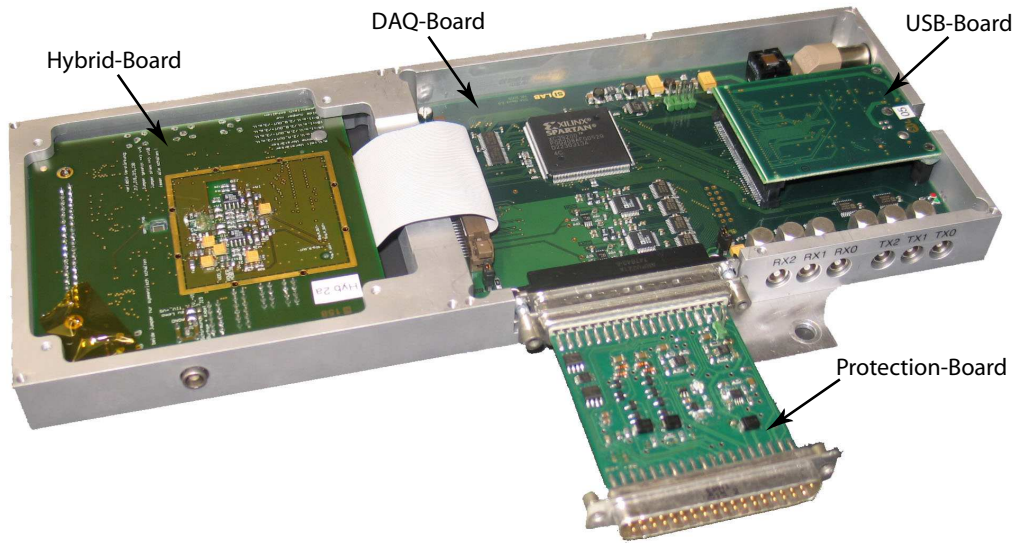}
  \caption{Photograph of the ILC DEPFET pixel system consisting of USB-Board, DAQ-Board and Hybrid-Board. The optional Protection-Board is attached at the bottom of the system.}
  \label{System_Photo}
\end{figure*}

As an intermediate step towards a full scale ILC detector ladder, a
prototype system using a \mbox{$128\times64$} DEPFET pixel matrix has been developed.
A schematic overview of the system is given in figure~\ref{System_overview}.
It consists of three major parts: a USB-Board, a DAQ-Board and a Hybrid-Board.
The USB-Board is based on USB\,2.0 standard and provides the communication between the system and a PC.
The Hybrid-Board hosts the DEPFET pixel matrix, the steering chips SWITCHER~II, the readout chip CURO~II and a pair of transimpedance amplifiers (I2U) that convert the current outputs of the CURO chip to voltage signals.
The DAQ-Board carries two 14\,bit ADCs for digitization and a SRAM for data storage.
Integral part of the DAQ-Board is a SPARTAN 3 FPGA which provides the configuration of all chips in the system and manages the synchronization between the components during data acquisition.
An external Protection-Board can be inserted to safeguard the power supply voltages.  
A photograph of the DEPFET system is shown in figure~\ref{System_Photo}.

\begin{figure}
  \includegraphics[width=80mm]{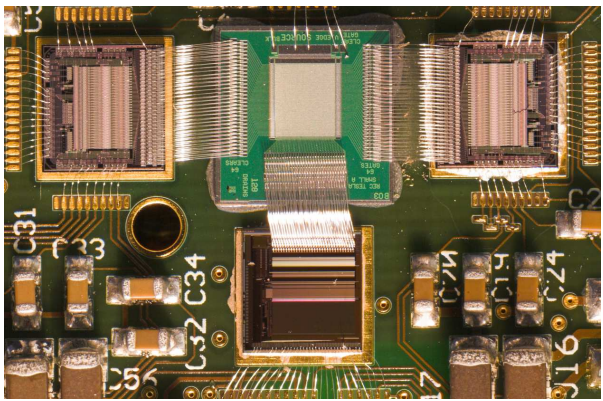}
  \caption{Chip assembly on the Hybrid-Board showing the DEPFET matrix in the middle, the two steering chips SWITCHER~II at both sides and the readout chip CURO~II at the bottom (area shown $\rm{\sim23\times15\,mm^2}$).}
  \label{Chip_Assembly}
\end{figure}

Figure~\ref{Chip_Assembly} shows a close-up view of the chip assembly on the Hybrid-Board.
The $128\times64$ pixel DEPFET matrix is situated in the center.
The steering strobes for the matrix are provided by two SWITCHER~II chips \cite{CITE_PHD_Peric} located at both matrix sides.
The SWITCHER~II has been fabricated in an AMS high voltage process to provide a voltage range of up to 25\,V, needed to operate some matrix designs.
The present hybrid uses two steering chips to be very versatile in operating different matrix types.
Recent measurements \cite{CITE_SDSSandow} show that for some matrix designs only one steering chip is sufficient to take care of matrix steering and clearing.

At the bottom of the matrix the 128 channel readout chip CURO~II \cite{CITE_PHD_Trimpl} is placed.
This chip is fabricated in a $0.25\,\mu$m TSMC process.
In addition, most of the design already uses radiation tolerant layout rules \cite{CITE_Snoeys}.
It is therefore very likely that the readout chip sustains the radiation dose of 200\,kRad expected after 5 years operation of the ILC.
Whether the present chip already shows a suitable radiation tolerance or if the layout has to be improved further on will be studied in irradiation test.

One of the main features of the readout chip are on-chip pedestal subtraction and zero suppression.
The zero suppression is performed by a scanner which scans the binary hit pattern in parallel.
Standalone tests with the present CURO~II chip show that the scanner finds up to two hits in a 128 channel hit pattern with rates of more than 100\,MHz.
This is much faster than needed to cope with the expected rates at the ILC. 
\section{Lab Measurements}
\begin{figure}
  \includegraphics[width=80mm]{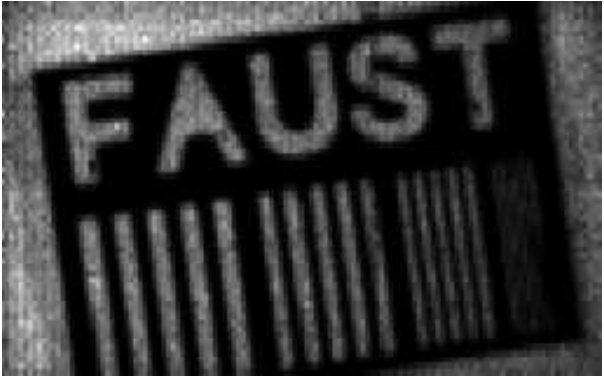}
  \caption{Radiogram of a $\rm{2\times3\,mm^2}$ tungsten test chart taken with $\rm{^{55}Fe}$.}
  \label{LAB_FAUSTLogo}
\end{figure}

The system has been used in the lab for X-ray detection in an energy range from 5.9 up to 44.23\,keV.
The capability of spatial detection of X-rays has been demonstrated using a 75\,$\mu$m thick tungsten test chart with an engraved logo and several line structures having a pitch of 100, 75, 50 and 25\,$\mu$m (from left to right, respectively).
The radiogram of the $\sim$2\,x\,3\,$\mathrm{mm^2}$ large test chart is shown in figure~\ref{LAB_FAUSTLogo}.
The test chart has been placed on the backside of the detector and the system has been irradiated with a $\mathrm{^{55}Fe}$ source.
The radiogram has been obtained by summing the pulse heights for pixels that contain a signal higher than five times their noise.
Although no spatial reconstruction techniques have been applied to improve the spatial resolution, the 50\,$\mu$m lines in the radiogram are clearly visible.
%
%

\begin{figure}
  \hskip-5mm
  \includegraphics[width=75mm]{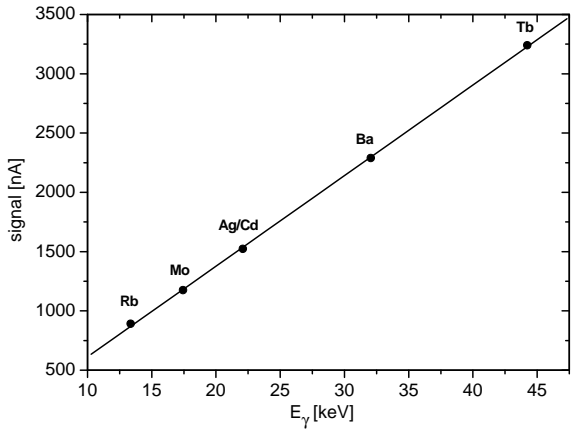}
  \caption{Gain measurement of the DEPFET pixel system. The sensor response is plotted for different radioactive sources.}
  \label{LAB_MatrixGain}
\end{figure}

To determine the internal gain of the DEPFET device and to measure the linearity of the system, energy spectra have been taken for Rubidium (Rb), Molybdenum (Mo), Silver (Ag), Cadmium (Cd), Barium (Ba) and Terbium (Tb), provided by a variable X-ray source \cite{CITE_Amersham}.
After performing a row-wise common mode correction\footnote{Subtraction of a mean value (common mode) from the pixel pedestal. The common mode is computed using the pedestals of each pixel of a row that contain no signal.}, clusters are reconstructed by identifying neighboring pixels, where every pixel contains a signal higher than five times its noise.
In figure~\ref{LAB_MatrixGain} the mean cluster signal as a function of the characteristic X-ray energy is shown.
From the slope of the system response an internal gain of the DEPFET sensor of $\mathrm{g_q = 282.6\pm3.3\,pA/e^-}$ is extracted.
The integral-non-linearity (INL) of the system is better than 0.8\,$\%$ for a dynamic range of 8500\,$\mathrm{e^-}$.
This range is large enough for the detection of about 2 MIPs in a 50\,$\mu$m thin sensor device.

The best noise performance achieved in the lab is around ENC=220\,e$^-$ at room temperature.
This number is quite comparable with the calculated noise performance
of $\rm{ENC\sim190\,e^-}$, where thermal noise is identified as the dominant contribution \cite{CITE_TrimplSDS}.
However, the total noise needs to be improved by a factor of two to achieve the final aim of around 100\,e$^-$.
This is within a realistic scope since the bandwidth of the present readout and steering chips have already been designed to cope with the final line rate of 20\,MHz.
It is therefore very likely that a future generation of the system with an optimized readout chip and sensors with higher $\rm{g_q}$ can achieve a noise of 100\,e$^-$.
\section{Beam Tests}
The system has been tested for MIP detection in the 6\,GeV electron test beam at the DESY synchrotron.
The analysis of the data is done in the following way:
After pedestal and common mode corrections, clusters are searched for.
Seeds are identified with a threshold cut of 5\,$\sigma$.
The seeds are combined with neighboring pixels if their signal is two times higher than the noise.

\begin{figure}
  \includegraphics[width=75mm]{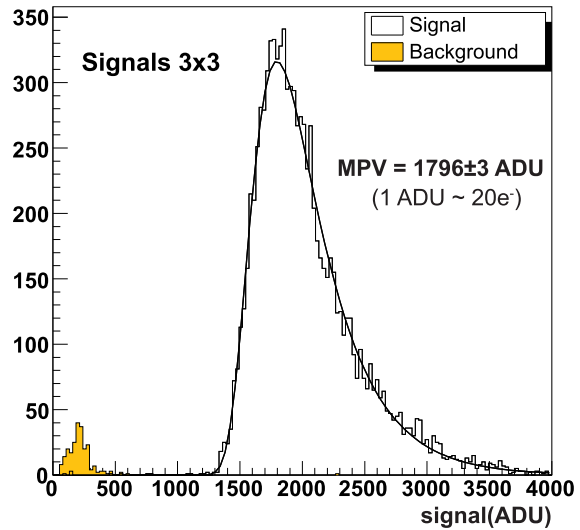}
  \caption{Signal distribution for clusters with up to $3\times3$ pixels.}
  \label{TB_signals}
\end{figure}

The signal distribution for clusters containing a maximum of $3\times3$ pixels is shown in figure~\ref{TB_signals}.
The observed most probable value of about 36000\,e$^-$ ($1796\pm3$\,ADU) obtained by a landau fit complies
with the expected MIP signal in a $450\,\mu$m silicon sensor material.
The system noise extracted from the noise peak of the raw data spectrum is found to be $15.723\pm0.007$\,ADU.
This combines to a signal-to-noise ratio of $114.2\pm0.2$.
The observed noise figure in the test beam ($\rm{\sim300\,e^-}$) is higher than the one achieved in the lab.
It turned out that the connection node between readout chip and transimpedance amplifier is very sensitive for pick-up.
Hence, the system noise depends on different operating parameters.
A future version of the readout chip implementing either a transimpedance amplifier or a current mode ADC is expected to overcome this.

\begin{figure}
  \hskip-5mm
  \includegraphics[width=73mm]{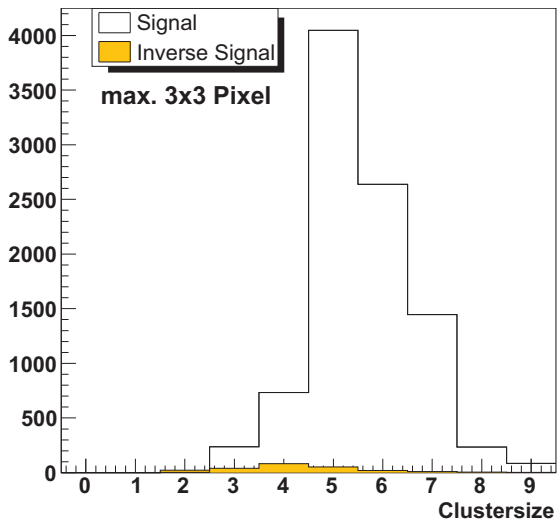}
  \caption{Cluster size distribution for a maximum of $3\times3$ pixels.}
  \label{TB_Clustersize}
\end{figure}

Figure~\ref{TB_Clustersize} shows the cluster size distribution.
The most probable cluster contains about 5 pixels only.
The relatively small clusters are due to the fact that charge collection in the DEPFET is done by an electric field and not by thermal diffusion.
Also shown in figure~\ref{TB_Clustersize} is the cluster size distribution for inverse signals.
This event distribution is obtained by inverting the signals and repeating the analysis.
It is very likely that the inverted signals are due to out-of-time events\footnote{Events that occur between the clear of the sensor and the pedestal
measurement.}.

\begin{figure}
  \hskip-5mm
  \includegraphics[width=74mm]{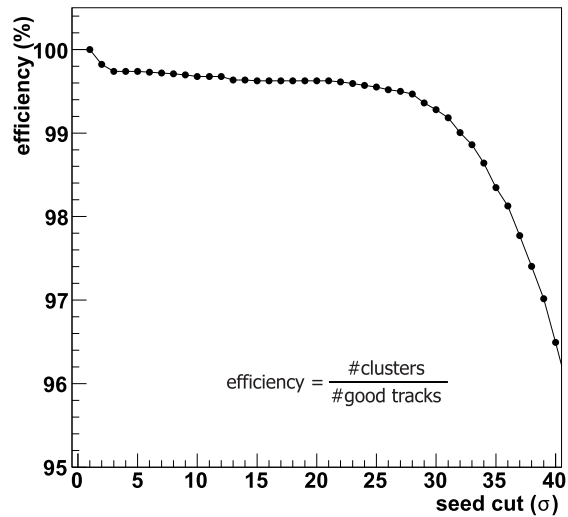}
  \caption{Efficiency to find a track as a function of the seed cut.}
  \label{TB_Efficiency}
\end{figure}

Furthermore, efficiency and purity for track reconstruction have been studied.
A four plane telescope system is used to determine the particle tracks. 
Clusters are searched in the DEPFET matrix within an area of $\pm2$\,pixels around the predicted track position.
The efficiency is defined as the ratio of clusters found in the fiducial DEPFET area and the corresponding tracks identified by the telescope system.
%
%
Figure~\ref{TB_Efficiency} shows the efficiency as a function of the seed cut threshold.
The observed inefficiency in the order of 0.3\,\% is mostly due to multiple scattering effects in the low energy electron beam. 
(Note, that clusters are detected in a relatively small area of $\pm$2 pixels.)
To reduce multiple scattering effects, only very stiff tracks are selected for the analysis by applying a high $\chi^2$ probability cut of 0.9995.
This enhances the efficiency to almost \mbox{100\,\% ($>99.99\,\%$)}.

\begin{figure}
  \hskip-5mm
  \includegraphics[width=75mm]{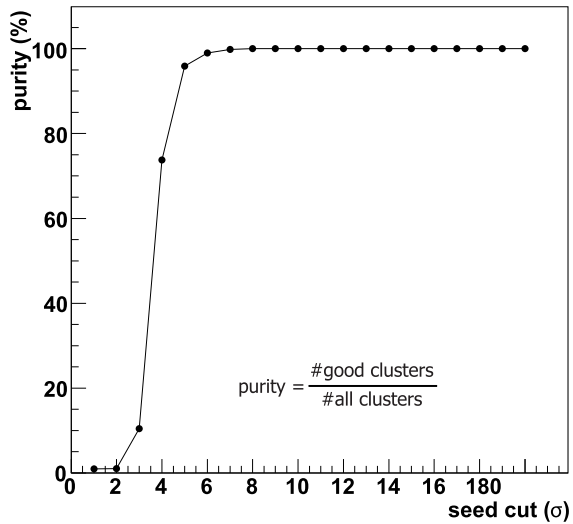}
  \caption{Track purity as a function of the seed cut.}
  \label{TB_Purity}
\end{figure}

The purity defined as the ratio of found good clusters and all clusters
%
%
is shown in figure~\ref{TB_Purity} as a function of the seed cut.
The main result of the efficiency and purity study is that an operation at 
almost 100\,\% purity and efficiency is achieved by choosing a seed cut of $5-7\,\sigma$.
Note, that this threshold corresponds to only $\sim5\,\%$ of a MIP signal.
\section{Conclusion and Outlook}
A prototype system for the ILC vertex detector based on DEPFET pixels has been developed.
The system has been successfully used for X-ray and MIP detection.
The current benchmarks of the system are:
\begin{itemize}
  \item{Noise performance of about ENC=220\,e$^-$,}
  \item{line rate of 2\,MHz,}
  \item{S/N$\sim$120 for MIPs using a $450\,\mu$m sensor.}
\end{itemize}
Concerning position resolution, the present test beam data yields an upper
limit of $5\,\mu$m dominated by multiple scattering effects.
A more precise measurement will be performed in the high energy test beam at CERN in summer 2006.

To achieve the final ILC specifications the system still needs improvements.
Most importantly, the line rate needs to be increased by a factor of 10.
The single components of the systems have already been tested up to a line rate of more than 20\,MHz,
whereas their synchronization in the system needs further optimization.
Moreover, the noise performance has to be improved by a factor of two.
Based on calculations and the experience with the present system,
we are very confident that this will be achieved with the next version of DEPFET sensors
and readout chip.

As a next step, new generations of all system components are designed and fabricated.
Large DEPFET matrices with up to $512\times512$ pixels are in production.
The pixel size is reduced to $\rm{25\times25\,\mu m^2}$ and the internal gain of the sensor is optimized.

A new steering chip using a standard, low voltage process is designed.
Some DEPFET designs having a so called \textit{cleargate structure} can be cleared at
much lower voltages than offered by the present steering chip.
Hence, the use of a submicron process is possible here as well.
This promises a much better radiation tolerance compared to the present chip using a high voltage technology.

A new generation of the readout chip is designed as well.
Operating the system with a duty cycle as close as possible to the bunch timing of the accelerator (1:199) will reduce the total power consumption significantly.
Hence, one of the most important conceptual extension of the readout chip will be a power down feature.

On a longer time scale, a major step is to combine the thinning technology with a DEPFET production in order to fabricate thin DEPFET sensors.
Together with the aimed noise figure of ENC=100\,e$^-$, a signal-to-noise ratio of 40 is expected for these thin devices.
%
%
\begin{acknowledgments}
The authors would like to thank the technology crew of PNSensor GmbH and the Semiconductor Laboratory at the MPI Munich supporting the DEPFET fabrication, and Uli K\"otz and Norbert Meyners for the invaluable support at DESY.
\end{acknowledgments}

\bigskip 

\begin{thebibliography}{99} 

\bibitem{CITE_DEPFET}
J. Kemmer, G. Lutz (1987): ``New semiconductor concepts'', NIM A253, pp.356ff.

\bibitem{CITE_DEPFET_Thinning}
L. Andricek et al. (2004): ``Processing of ultra thin silicon sensors for future linear collider experiments'', IEEE TNS, vol.51 No.3, pp.1117-1120

\bibitem{CITE_LaciRadHard}
L. Andricek et al. (2006): ``The MOS-type DEPFET Pixel Sensor for the ILC environment'', NIM A565, pp.165-171

\bibitem{CITE_Trimpl_VTX02}
M. Trimpl et al. (2003): ``A Fast Readout using Switched Current Techniques for a DEPFET-Pixel Vertex Detector at TESLA'', NIM A511, pp.257-264

\bibitem{CITE_SICell}
J.B. Hughes, N.C. Bird, I.C. Macbeth (1989): ``Switched Currents - A New Technique for Analog Sampled-Data Signal Processing'', IEEE ISCAS (217), pp.1584-1587

\bibitem{CITE_PHD_Peric}
I. Peric (2004): ``Design and Realization of Integrated Circuits for the Readout of Pixel Sensors in High-Energy Physics and Biomedical Imaging'', PhD thesis BONN-IR-2004-13, Bonn University

\bibitem{CITE_SDSSandow}
C. Sandow et al.: ``Clear-performance of linear DEPFET devices'', proceedings of the SDS 2005 conference to be published in NIMA

\bibitem{CITE_PHD_Trimpl}
M. Trimpl (2005): ``Design of a current based readout chip and development of a DEPFET pixel prototype system for the ILC vertex detector'', PhD thesis BONN-IR-2005-08, Bonn University, \textit{http://hss.ulb.uni-bonn.de/diss\_online}

\bibitem{CITE_Snoeys}
W. Snoeys et al. (2000): ``Layout techniques to enhance the radiation tolerance of standard CMOS technologies demonstrated on a pixel detector readout chip'', NIM A439, pp.349-360

\bibitem{CITE_TrimplSDS}
M. Trimpl et al.: ``Performance of a DEPFET pixel system for particle detection'', proceedings of the SDS 2005 conference to be published in NIMA

\bibitem{CITE_Amersham}
Amersham, AMC~2084 

\end{thebibliography}

\end{document}